\documentclass[%
 reprint,
 amsmath,amssymb,
 aip,
]{revtex4-2}

\usepackage[T1]{fontenc}
\usepackage{upgreek}

\usepackage{multirow}
\usepackage{graphicx}
\usepackage{dcolumn}
\usepackage{bm}
\usepackage{subcaption}
\usepackage{siunitx}
\usepackage{pifont}
\newcommand{\cmark}{\ding{51}}%
\newcommand{\xmark}{\ding{55}}%

\DeclareMathOperator*{\argmin}{arg\,min}

\makeatletter
\def\@email#1#2{%
 \endgroup
 \patchcmd{\titleblock@produce}
  {\frontmatter@RRAPformat}
  {\frontmatter@RRAPformat{\produce@RRAP{*#1\href{mailto:#2}{#2}}}\frontmatter@RRAPformat}
  {}{}
}%
\makeatother

\begin{document}

\title{Harnessing Data Augmentation to Quantify Uncertainty in the Early Estimation of Single-Photon Source Quality}

\author{David Jacob Kedziora}
\email{david.kedziora@uts.edu.au}
\affiliation{Complex Adaptive Systems Lab, University of Technology Sydney, Sydney, Australia}

\author{Anna Musiał}
\affiliation{Department of Experimental Physics, Wroclaw University of Science and Technology, Wrocław, Poland}

\author{Wojciech Rudno-Rudziński}
\affiliation{Department of Experimental Physics, Wroclaw University of Science and Technology, Wrocław, Poland}

\author{Bogdan Gabrys}
\affiliation{Complex Adaptive Systems Lab, University of Technology Sydney, Sydney, Australia}

\date{\today}

\begin{abstract}
Novel methods for rapidly estimating single-photon source (SPS) quality have been promoted in recent literature to address the expensive and time-consuming nature of experimental validation via intensity interferometry.
However, the frequent lack of uncertainty discussions and reproducible details raises concerns about their reliability.
This study investigates the use of data augmentation, a machine learning technique, to supplement experimental data with bootstrapped samples and quantify the uncertainty of such estimates. Eight datasets obtained from measurements involving a single InGaAs/GaAs epitaxial quantum dot serve as a proof-of-principle example.
Analysis of one of the SPS quality metrics derived from efficient histogram fitting of the synthetic samples, i.e.~the probability of multi-photon emission events, reveals significant uncertainty contributed by stochastic variability in the Poisson processes that describe detection rates.
Ignoring this source of error risks severe overconfidence in both early quality estimates and claims for state-of-the-art SPS devices.
Additionally, this study finds that standard least-squares fitting is comparable to using a Poisson likelihood, and expanding averages show some promise for early estimation. Also, reducing background counts improves fitting accuracy but does not address the Poisson-process variability.
Ultimately, data augmentation demonstrates its value in supplementing physical experiments; its benefit here is to emphasise the need for a cautious assessment of SPS quality.
\end{abstract}

\keywords{quantum dots; quantum communication; emission statistics; bootstrapping; generative models; uncertainty analysis}

\maketitle

\section{Introduction}
\label{Sec:Intro}

The world is presently witnessing the advent of the `Quantum 2.0' technological revolution, an era distinguished by the development of devices that, in their operation, involve manipulating quantum states of light and matter~\cite{OIDA}.
Indeed, these devices seek to exploit purely quantum phenomena, such as superposition or entanglement.
Unsurprisingly, many research fields are linked to this revolution, e.g.~quantum communication networks and cryptography~\cite{Sasaki2011, Liao2017, Schimpf2021}, quantum computing and simulation~\cite{computing, Deutsch2020, Zhong2020}, quantum imaging~\cite{Lemos2014, Tenne2018, Casacio2021}, quantum sensing within magnetic~\cite{Segawa2023, Tong2023} and gravitational~\cite{Mnoret2018, Stray2022} fields, and so on.
Likewise, considerably many engineering designs, both proposed and proven, take advantage of quantum optical phenomena, e.g.~by using photonic degrees of freedom as an information carrier, subject to transmission/manipulation, or to probe the state of a system.
However, to really exploit the quantum nature of light, non-classical light sources, e.g.~a single-photon source (SPS), are required.

Certainly, an SPS constitutes a valuable resource for both fundamental studies in quantum optics and industrial applications of quantum technology.
However, engineering an SPS is complicated as it must fulfil the exacting requirements of practicality, e.g.~a feasibly attainable operating temperature, as well as a desired emission wavelength, emission rate and collection efficiency.
More advanced application schemes may even require polarisation control or indistinguishability between consecutive photons.
Nonetheless, the most crucial characteristic of an intended SPS is the purity of single-photon emission. While other secondary attributes are important, they become irrelevant if the essential criterion for an SPS is not fulfilled. This purity is typically measured by a photon-intensity autocorrelation experiment, which determines the second-order auto-correlation function $g^{(2)}(\tau)$~\cite{Kimble1977, Kurtsiefer2000, Zwiller2001, Buckley2012}.

Notably, many material systems have been proposed for the engineering of an SPS, starting with systems based on atoms and ions.
These are clean and reproducible, but require bulky and cumbersome setups, limiting their practical applications~\cite{Kuhn2002, Blinov2004}.
In contrast, solid-state systems appear to have better long-term prospects, and several classes of related structures show promise as single-photon emitters: quantum dots grown by colloidal~\cite{Chandrasekaran2017, Zhu2022} or epitaxial techniques~\cite{Yuan2002, 2017, Schimpf2021}, as well as various defect centres, e.g. in diamonds~\cite{Doherty2013, Hensen2015}, silicon carbide~\cite{Bathen2021}, hexagonal boron nitride~\cite{Sajid2020}, and 2D materials~\cite{He2015, Gao2023}.

Whatever the practical realisation of an SPS is, the quality of the final device must be precisely assessed. This paper will focus solely on the purity of single-photon emission.
Ideally, an SPS works as follows: a quantum emitter with a discrete energy spectrum is repeatedly excited optically or electrically, each time with relaxation following excitation, and thus emits a train of photons. Every excitation pulse it encounters should result in the emission of just a single photon, i.e.~it must act as an on-demand source of single photons. Proving that an object is classifiable as an SPS typically requires using an intensity-based `Hanbury Brown and Twiss' interferometer~\cite{brtw56} to experimentally determine $g^{(2)}(\tau)$. Specifically, the emission is split equally between two detection paths. A detector with single-photon sensitivity is placed at the end of each path. Dedicated electronics then build up a histogram of two-photon events as a function of the temporal separation between the output signals from the two detectors. These events occur when both detectors are separately triggered. Any events at an effective time delay of zero imply the SPS underwent multi-photon emission (MPE), i.e.~$g^{(2)}(0)>0$. Indeed, the quality of an SPS depends on how often an MPE event triggers a two-photon detection. Excessive MPE is undesirable as it limits the utility of an intended SPS, e.g.~causing errors in quantum communication and cryptography protocols~\cite{Waks2002}.

The problem with determining SPS quality is that the measurement is statistical in nature and, therefore, a sufficiently large amount of data needs to be collected to derive reliable conclusions.
This accumulation can take substantial time, even on the order of days, if the probability of observing two-photon events is low.
Likewise, measurements can be resource intensive, given that single-photon emission is typically observed at expensive-to-maintain cryogenic temperatures.
Specifically, the most efficient single-photon detectors need to be kept below the critical temperature of a superconducting transition, while the highest-quality SPS devices, i.e.~III-V semiconductor epitaxial quantum dots, require cryogenic temperatures for optimal performance.
Then there is the aim of the measurement. For proof-of-principle experiments, obtaining a value of $g^{(2)}(0)<0.5$ will be sufficient, while, for state-of-the-art engineering, ensuring a minimal $g^{(2)}(0)$ is critical~\cite{mita16, hafi18, scjo18}.
Obtaining statistical significance in the latter case can take a substantially longer measurement.
The situation worsens when working with more complex characteristics, e.g.~photon indistinguishability or the degree of quantum entanglement, for which quantum state tomography requires many cross-correlation measurements to be performed in different bases~\cite{James2001, Huber2017}.
In short, physicists in the field keenly await any improvements in data acquisition or analysis techniques that might give accurate SPS quality estimates for far fewer two-photon event detections, i.e.~significantly reduced experiment times.

Given this context, recent literature, such as an article proposing a maximum \textit{a posteriori} (MAP) scheme~\cite{coad20}, has drawn much attention, suggesting that small data samples are sufficient to provide reasonable estimates of SPS quality in terms of MPE probability, i.e.~$g^{(2)}(0)$.
This work investigates how true such claims are.
Exemplary experimental datasets are readily available for such an analysis, i.e.~measurements for epitaxial InGaAs/GaAs quantum dots within deterministic photonic nanostructures emitting at 1.3 \si{\um}~\cite{muzo20}, where a 10 s collection interval for two-photon events provides insight into how the correlation histogram can build up over time.
It is then not difficult to examine various statistical fitting methods, assessing how estimates of SPS quality and their fitting-based accuracy develop with respect to the growing number of detected two-photon events.
However, herein lies a major weakness.
Who is to say that one sequence of observations, already very costly, is a typical reflection of how the histogram and associated quality estimates evolve?
How can one even be sure that the final long-time estimate, used as a baseline of comparison for a short-time estimate, is itself accurate?

In data science and machine learning, when information is sparse, generative procedures are sometimes used to simulate new data from existing datasets in a process often called `data augmentation'~\cite{hama20, mamo22}.
This work continues the trend of applying data-science principles to experimental physics~\cite{jaon20, tawu21} by leaning on a technique that is effectively `bootstrapping'. Each experimental dataset is used to generate arbitrarily many alternative samples that then determine the sampling-based variance of an SPS-quality estimand.
After highlighting the non-negligible error bars on even late-stage estimates of $g^{(2)}(0)$, this article reinforces a cautionary message when evaluating new approaches for fast estimation: \textit{do not underestimate stochastic variability}.

This work is organised and presented in the following manner.
Section~\ref{Sec:Theory} provides, with a novel reformulation suited for efficient computation, the theoretical equations that describe the expected structure of a two-photon event histogram.
Section~\ref{Sec:Methodology} describes the eight datasets analysed in this work (\ref{Sec:Datasets}), the two investigated methods of fitting data for SPS quality estimation (\ref{Sec:Fitting}), and the procedure for generating synthetic data in support of calculating error bounds (\ref{Sec:Augmentation}).
Section~\ref{Sec:Results} then presents the results of numerically fitting theoretical equations to the data, both observed and synthetic, assessing (1) the expected sample-based errors of $g^{(2)}(0)$, (2) whether a MAP scheme~\cite{coad20} is superior to least-squares histogram fitting, (3) the possibility of using expanding averages for early $g^{(2)}(0)$ estimation, and (4) the impact of background counts on accuracy.
Section~\ref{Sec:Discussion} subsequently takes the lessons learned here to critically reexamine and discuss the article that motivated this work~\cite{coad20}.
Finally, Section~\ref{Sec:Conclusion} summarises the conclusions of this research.

\section{Theory}  
\label{Sec:Theory}

SPS quality estimation typically leverages the technique of intensity interferometry with single-photon detectors~\cite{brtw56}, albeit ones that cannot resolve the number of photons. Because of this limitation, the photon flux must be kept sufficiently low for the measured histogram to reflect actual photon statistics.
Thus, in many cases, only a handful of two-photon events is detected, per second, within a temporal-separation window of interest. 
Consequently, in ideal operating conditions, these independent events can be modelled as elements of Poisson processes.
Specifically, given any particular temporal separation $\tau$ between the two photons incident on the two detectors, the probability of encountering $n$ such events after $t$ seconds is
\begin{equation}
\label{Eq:Poisson}
    Pr(n;R,t) = \frac{(Rt)^n}{n!} e^{-Rt},
\end{equation}
where $R$ is a mean detection rate in Hz.
This rate technically depends on how $\tau$ values are binned, but matters of discretisation are ignored in this section; they are trivial to apply during numerical implementation.

In theory, because the mean detection rate $R$ is fixed for any given $\tau$, an accumulated histogram of detected events should closely approximate $R(\tau)\times t$ in the long-time limit of $t$.
Moreover, the expected form of $R(\tau)$ is well-known in many cases.
However, before defining this function, it is essential to note that many experimental datasets report two-photon event delays in raw uncalibrated fashion, $\tau_r$, so an offset that aligns MPE events with a time separation of zero must often be determined, i.e.~$\tau = \tau_r - \tau_0$. 
Part of the offset is due to the difference in length between optical paths from the beam splitter to the detectors, as well as the length of the electrical connections. It is constant for a given experimental configuration. Additionally, as the triggering order of detectors determines the sign of a time delay, an electronic offset applied during measurements will also typically be present, thus capturing a portion of the negative time delays.
Now, in the case of SPS emission driven by a pulsed laser with period $\Lambda$, detected two-photon events should peak in number for temporal separations of $k\Lambda$, where $k\neq 0$.
This outcome represents the detector identifying pairs of photons released by temporally neighbouring laser pulses.
In scenarios where bunching on long time scales is observed, the envelope of these peaks can be considered to decay over extended ranges of time delays, i.e.~by a factor of $\gamma_e$.
As for the individual peaks, their spread across $\tau$ values can be modelled as two-sided exponential functions with decay factor $\gamma_p$.

Putting this all together, the mean rate of observing two-photon events with time separation $\tau$ is
\begin{equation}
\label{Eq:FitOld}
    R(\tau; \theta) = R_b + R_p e^{-\gamma_e |\tau|} \left( g e^{-\gamma_p |\tau|} + \sum_{k\neq 0} e^{-\gamma_p |\tau-k\Lambda|}\right),
\end{equation}
where $R_b$ represents background detections predominantly caused by detector dark counts, $R_p$ denotes the peak rate of pulse-driven events, and $g$ defines the ratio of MPE events to non-MPE events.
Overall, the function has seven parameters when compared against experimental data defined over $\tau_r$, i.e.~$\theta = \left\{\tau_0, R_b, R_p, g, \gamma_e, \gamma_p, \Lambda\right\}$.
In particular, $g$ is one of the indicators of SPS quality, and it is often written as $g^{(2)}(0)$ to denote second-order coherence/correlation at zero time delay; the short form will be used in the rest of the text for convenience.
A value of $g=0$ indicates a perfect SPS.

Now, much of the work in this article involves fitting Eq.~(\ref{Eq:FitOld}) to experimental observations of two-photon events and their time delays.
However, infinite sums, even finitely truncated, are not computationally efficient to calculate.
So, using the floor function to apply a modulo operation, i.e.~$\tau_m = \textrm{mod}(\tau,\Lambda) = \tau - \Lambda \left\lfloor\tau/\Lambda\right\rfloor$, a useful insight is as follows:
\begin{alignat}{3}
\label{Eq:Conversion1}
    \nonumber \sum_{k\in \mathbb{Z}} e^{-\gamma_p |\tau-k\Lambda|} &{}={}&& \sum_{k\in \mathbb{Z}} e^{-\gamma_p |\tau_m-k\Lambda|}, \\
    \nonumber &{}={}&& e^{-\gamma_p\tau_m}\sum_{k\in \mathbb{Z}_0^+}e^{-k\gamma_p\Lambda} \\
    &&&+ e^{\gamma_p(\tau_m-\Lambda)}\sum_{k\in \mathbb{Z}_0^+}e^{-k\gamma_p\Lambda}.
\end{alignat}
Effectively, for any value of $\tau$, this term has the same answer as it does for some $\tau$ between $0$ and $\Lambda$, i.e.~an infinite sum of concave-right exponentials to the left and concave-left exponentials to the right.
Here, a particular equality proves useful: $1+e^{-x}+e^{-2x}+\ldots=e^x/(e^x-1)$.
Hence, defining $T=\gamma_p\tau_m$ and $L=\gamma_p\Lambda$ for convenience, Eq.~(\ref{Eq:Conversion1}) can be further manipulated as follows:
\begin{alignat}{2}
\label{Eq:Conversion2}
    \nonumber \left(e^{-T}+e^{(T-L)}\right)\left(\frac{e^L}{e^L-1}\right) &= 
    \frac{e^{\left(-T+\frac{L}{2}\right)}+e^{\left(T-\frac{L}{2}\right)}}{e^\frac{L}{2}-e^{-\frac{L}{2}}} \\
    &= \frac{\cosh{\left(T-\frac{L}{2}\right)}}{\sinh{\left(\frac{L}{2}\right)}}.
\end{alignat}

Thus, by trading an infinite series for reliance on the modulo function, Eq.~(\ref{Eq:FitOld}) can be written as a much more computationally amenable expression:
\begin{align}
\label{Eq:FitNew}
    \nonumber R(\tau; \theta) {}={}& R_b + R_p e^{-\gamma_e |\tau|} \\
    &\times\left( \left(g-1\right) e^{-\gamma_p |\tau|} + \frac{\cosh{\left(\gamma_p\left(\tau_m-\frac{\Lambda}{2}\right)\right)}}{\sinh{\left(\frac{\gamma_p\Lambda}{2}\right)}}\right).
\end{align}
This form of the equation makes it explicit that the mean rate for detected two-photon events appears as a repeating $\cosh$ function over the domain of $\tau$, albeit with the peak around $\tau=0$ diminished according to SPS quality; this is the peak that represents MPE events.

\section{Methodology}
\label{Sec:Methodology}

The research described in this article revolves around the computational analysis of experimental interferometry measurements. To support reproducibility, the data and Python scripts are available at https://github.com/UTS-CASLab/sps-quality, commit f1782ff. 
Here, the experimental datasets are detailed in Sec.~\ref{Sec:Datasets}, two fitting methods for estimating $g$ are described in Sec.~\ref{Sec:Fitting}, and the procedure for augmenting the experimental data, i.e.~generating synthetic samples of observations, is elaborated in Sec.~\ref{Sec:Augmentation}.

\subsection{The Datasets}
\label{Sec:Datasets}

This work selects eight experimental datasets for investigation, all sourced from prior attempts to 
engineer/assess a `fibre-coupled semiconductor single-photon source for secure quantum-communication in the 1.3 \si{\um} range' (FI-SEQUR).
Specifically, they all involve the same transition for the same single InGaAs/GaAs epitaxial quantum dot, which is positioned deterministically with respect to the centre of a photonic mesa structure. 
The growth and deterministic fabrication methods~\cite{sro18, zo19}, as well as the optical setup for assessing SPS quality~\cite{muzo20}, are detailed elsewhere.

\begin{table}[htb!]
\centering
\caption{Descriptive summary of the eight FI-SEQUR datasets.}
\label{Tab:Datasets}
\begin{tabular}{|c|ccccc|}
\hline
\begin{tabular}[c]{@{}c@{}}Shorthand\\ Label\end{tabular} &
  \begin{tabular}[c]{@{}c@{}}Laser\\ Intensity\\ (\si{\uW})\end{tabular} &
  \begin{tabular}[c]{@{}c@{}}Single\\ Detector\\ Counts\\ Per Second\\ (Hz)\end{tabular} &
  \begin{tabular}[c]{@{}c@{}}Total\\ Events\end{tabular} &
  \begin{tabular}[c]{@{}c@{}}Total\\ Duration\\ (s)\end{tabular} &
  \begin{tabular}[c]{@{}c@{}}Average\\ Event\\ Rate\\ (Hz)\end{tabular} \\ \hline
1p2uW & 1.2 & 3000  & 65037 & 23950 & 2.716  \\
2p5uW & 2.5 & 4000  & 51534 & 10510 & 4.903  \\
4uW   & 4   & 4100  & 61246 & 9390  & 6.522  \\
8uW   & 8   & 5100  & 21611 & 2920  & 7.401  \\
10uW-- & 10  & 6000  & 28817 & 3010  & 9.574  \\
10uW+ & 10  & 12000 & 45198 & 1210  & 37.354 \\
20uW  & 20  & 7000  & 55469 & 4780  & 11.604 \\
30uW  & 30  & 7000  & 52088 & 4780  & 10.897 \\ \hline
\end{tabular}
\end{table}

Each dataset, summarised in Table~\ref{Tab:Datasets}, results from a separate experiment where an 80 MHz laser of a fixed output power excites the SPS into emitting a train of photons.
Here, the excitation is above-band, with the semiconductor laser operating at a wavelength of 805 nm and a pulse length of 50 ps.
In contrast, many SPS devices reaching state-of-the-art $g^{(2)}(0)$ values~\cite{mita16, hafi18, scjo18} rely on coherent excitation schemes.
Their trade-off is that the source laser tends to be more complex, both in physical implementation and size, and therefore less practical, e.g.~requiring precise wavelength tuning for each quantum dot and a heavy spectral filtering of laser photons that reduces source brightness.
However, excitation details are irrelevant to the data augmentation/analysis technique in this paper; only the obtained histogram matters.

After emission, the actual number of photons received by each detector is then measured in counts per second (CPS).
This value depends on the efficiency and associated losses inherent within the experimental setup, e.g.~the quantum efficiency of the single-photon detectors, as well as the source itself.
On the source side, three factors are important: (1) the probability of occupation for the emitting state that is driven by the excitation power, (2) internal quantum efficiency, i.e.~the probability that the excited quantum dot emits a photon rather than non-radiatively relaxing, and (3) collection efficiency, i.e.~the percentage of emitted photons that can be collected by the detection optics.
Only the first factor should vary between the investigated datasets.
Specifically, increased excitation power means a higher number of carriers are available to be captured by a quantum dot, resulting in shorter periods of time when the quantum dot is unoccupied.
Photons cannot be generated without this occupation.
However, two of the datasets have the same excitation power and different CPS values.
This variation is possibly due to de-adjustment within the experimental setup, e.g.~from temperature fluctuations in the lab.

Regarding contents, each dataset is a 2D matrix of two-photon event counts, binned in 10-second detection `snapshots' along one axis and intervals of time separation along the other.
Specifically, the delay domain of interest ranges from $0$ ns to approximately $500$ ns, with each bin typically covering a $\Delta\tau$ of $0.256$ ns; it is $0.128$ ns for the 1.2 \si{\uW} experiment.
Importantly, these are raw $\tau_r$ values, and the actual zero delay, i.e.~MPE, occurs at about $60$ ns due to the electronic offset discussed previously.

\begin{figure}[htb!]
\centering
\includegraphics[width=\columnwidth]{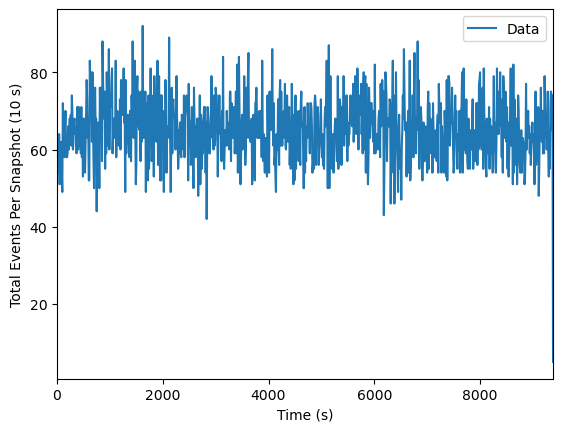}
\caption{\label{Fig:ExampleSignal} The total number of two-photon events ($0\leq \tau_r\lesssim \num{5e-7}$) detected within every $10$ s snapshot for the duration of the 4uW experiment.}
\end{figure}

\begin{figure}[htb!]
\centering
\begin{subfigure}{\columnwidth}
\centering
\includegraphics[width=\columnwidth]{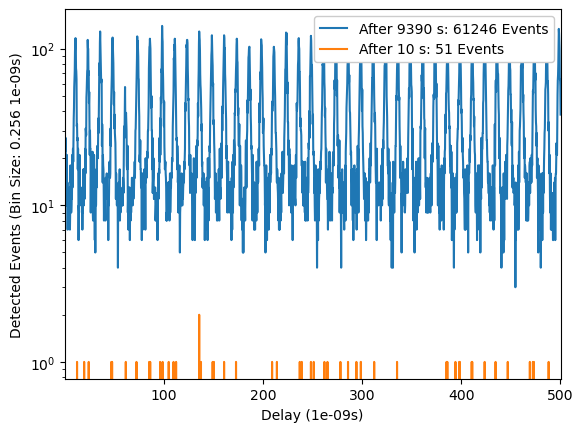}
\caption{}
\label{Fig:ExampleHist}
\end{subfigure}
\begin{subfigure}{\columnwidth}
\centering
\includegraphics[width=\columnwidth]{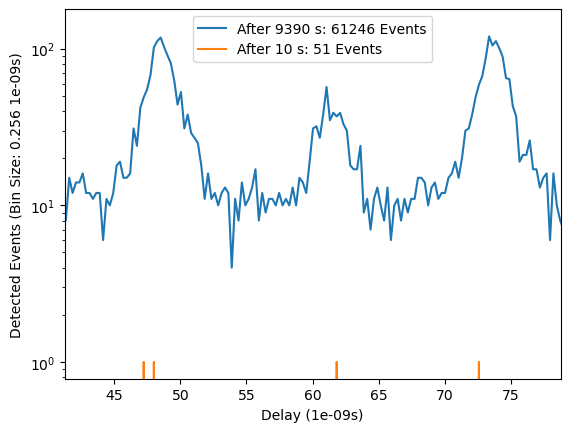}
\caption{}
\label{Fig:ExampleHistCloseup}
\end{subfigure}
\caption{Histogram of two-photon events detected during the entire 4uW experiment, as well as during the first $10$ s snapshot. The count axis is logarithmic. (a) A view of the entire $\tau_r$ domain. (b) A closeup around the MPE peak.}
\end{figure}

Notably, the dataset matrix can be summed along one axis to display how many two-photon events within the delay domain of interest are detected over time.
Such a summation also enables calculating an average event rate within this domain.
It is routine and unremarkable for each event rate to be far lower than its corresponding CPS value; encountering a \textit{pair} of photons across detectors with sub-microsecond temporal separation is relatively rare.
An example signal constructed from the 4uW experiment is displayed in Fig.~\ref{Fig:ExampleSignal}, where the visually discernible mean of approximately $65$ events per $10$ s snapshot corresponds with the average event rate of $6.522$ Hz listed in Table~\ref{Tab:Datasets}.
However, the histogram compiled along the other axis of the dataset, i.e. the $\tau_r$ domain, is of greater interest.
As Fig.~\ref {Fig:ExampleHist} shows, one snapshot of $10$ s provides a limited amount of information.
In contrast, if the events from all snapshots are combined, a 40-peak `comb' structure appears in the histogram.
One of these peaks is smaller than all others, clarified by the Fig.~\ref{Fig:ExampleHistCloseup} closeup, and, after accounting for the background, the relative amplitude of this MPE peak characterises the quality of an SPS.

\subsection{The Fitting Procedure}
\label{Sec:Fitting}

Core to this research is the notion of fitting Eq.~\ref{Eq:FitNew} to histograms of data.
Specifically, this work uses the Python LMFIT package~\cite{nest14} to, by default, minimise a sum of squared errors, thus obtaining the following optimised `least-squares' (LS) parameters:
\begin{equation}
\label{Eq:ObjectiveLS}
    \theta_{LS} = \argmin_\theta \sum_i \left(R(\tau_i; \theta)t - d_i\right)^2,
\end{equation}
where $i$ is the index of a bin centred at $\tau_i$ -- this implicitly redefines $R$ as a discrete function -- and $d_i$ is the number of events detected in that bin after $t$ seconds.
However, it has also been argued that there may be a better way than least-squares fitting.
In short, an MAP proposal~\cite{coad20} suggests using the following `Poisson-likelihood' (P) objective function instead:
\begin{equation}
\label{Eq:ObjectiveP}
    \theta_{P} = \argmin_\theta \sum_i \left(R(\tau_i; \theta)t - d_i\ln\!\left(R(\tau_i; \theta)t\right)\right).
\end{equation}
Conveniently, by way of defining a \texttt{reduce\_fcn} argument, the LMFIT package allows fitting procedures to minimise either of the two objectives described by Eq.~(\ref{Eq:ObjectiveLS}) and Eq.~(\ref{Eq:ObjectiveP}) without affecting other processes, e.g.~the automatic calculation of standard errors.

\begin{table}[htb!]
\centering
\caption{Initial values and parameter constraints used when fitting the mean-rate function in Eq.~(\ref{Eq:FitNew}), times duration $t$, to the data $d_i$ accumulated over that period.}
\label{Tab:Bounds}
\begin{tabular}{|c|cccc|}
\hline
Parameter       & Vary   & Min             & \begin{tabular}[c]{@{}c@{}}Initial Value\end{tabular} & Max      \\ \hline
$\tau_0$ (s)        & \cmark & \num{55e-9}           & \num{60e-9}                                                   & \num{65e-9}    \\
$R_b$ (Hz)           & \cmark & 0               & $\left(\sum_i d_i/t\right)/\left(\num{5e-7}/\Delta\tau\right)$                                       & $\infty$ \\
$R_p$ (Hz)           & \cmark & 0               & $\left(\sum_i d_i/t\right)/40$                                         & $\infty$ \\
$g$             & \cmark & 0               & 0.5                                                     & 1        \\
$\gamma_e$ (s\textsuperscript{$-1$})      & \xmark & -               & 0                                                       & -        \\
$1/\gamma_p$ (s) & \cmark & $\Delta\tau/10$ & $\Delta\tau$                                            & $\infty$ \\
$\Lambda$ (s)       & \xmark & -               & \num{1.25e-8}                                                 & -        \\ \hline
\end{tabular}
\end{table}

To maintain consistency with the MAP proposal~\cite{coad20}, this work employs the Powell method for the fitting procedure, although paired with a subsequent Trust Region Reflective optimisation (\texttt{method="least\_squares"}).
For completeness, the optimisation bounds and initial values for the parameters are listed in Table~\ref{Tab:Bounds}, noting that pulse period $\Lambda$ is experimentally fixed and $\gamma_e$, the envelope decay, is assumed to be negligible.
Given these settings, a single-threaded completion of the five-parameter fitting procedure generally takes a second or less on a 3.50 GHz Intel Core i9-9900X CPU with 32 GB of RAM.
Such speed allows histogram fitting to be part of any online method for estimating $g$, easily running alongside an actual event-detection experiment.

\subsection{Data Augmentation}
\label{Sec:Augmentation}

Each of the eight datasets in Table~\ref{Tab:Datasets} represents a resource-expensive run of over 20 minutes, usually several hours, and each experiment observes between 20000 and 70000 two-photon events of interest in that time, i.e.~with $0\leq \tau_r\lesssim \num{5e-7}$.
In theory, histograms compiled from the maximum number of events observed are likely to provide the best estimates of quality parameter $g$.
A naive approach then would be to engineer a novel fitting algorithm and assess its virtues simply by how quickly its estimate of $g$ -- the fewer two-photon events in a histogram, the better -- lands within fitting-based error bars of the maximum-time $g$ estimate.
However, the problem with this is as obvious as it is easily overlooked.
What guarantee is there that the evolution of a second-order auto-correlation function $g^{(2)}(\tau)$, as exhibited by any one experiment, is typical?
It would be disingenuous to hype up an estimation approach based on the off-chance that early detections of coincidences occur in just the right pattern.
Then there is a more conservative question: are even $45000\pm25000$ observations sufficient to treat derived $g^{(2)}(0)$ values as ground truths?

One way to address these questions is to synthetically generate new data based on the existing datasets, i.e.~engage in data augmentation~\cite{hama20, mamo22}.
Early versions of the research in this article considered doing so by simply shuffling the $10$ s snapshots available, assuming them to be independent.
However, with such a method, the benefit of statistical independence is gradually lost over the summation of snapshots.
Instead, the approach selected for this work leans on the accuracy of the assumptions introduced in Sec.~\ref{Sec:Theory}, namely the Poisson statistics governing two-photon event detections.
Expressly, per dataset, assume the optimised fit of Eq.~(\ref{Eq:FitNew}) applied to all observations is the ground-truth mean rate of detections, i.e.~fit $R(\tau_i; \theta_{\textrm{best}}) \times t_{\textrm{total}}$.
It is then possible to generate new data based on these ground truths as follows.
For a given duration $t_{\textrm{new}}$ and a temporal-separation bin centred at $\tau_i$, randomly sample from a Poisson distribution with mean-value parameter $R(\tau_i; \theta_{\textrm{best}}) \times t_{\textrm{new}}$.
Then, repeat this process for every bin index $i$ until a new histogram is generated across the entire delay domain.

The main benefits of this technique, reminiscent of bootstrapping approaches used in the field~\cite{bama21, alku22}, are that (1) the new samples are independent, supporting statistical rigour, (2) thousands of synthetic datasets can be generated computationally at speed, (3) the method can extrapolate real-world stochastic data to simulate what is out of feasible reach for a typical experiment, e.g.~the accumulation of a million observations, and (4) subsequent fits of the synthetic histograms indicate expected sampling-based errors for quality parameter $g$.
Best of all, this procedure is generally applicable, provided that (1) two-photon event detection adheres to Poisson statistics and (2) the long-term shape of a two-photon event histogram can be theoretically described for an emission context. Thus, although eight variations involving a single quantum dot are used in this paper to showcase the technique, hence enabling convenient comparative analysis with minimal conflating factors, it can just as easily be applied to an SPS stimulated by a continuous-wave laser, a thermal light source~\cite{coad20}, and so on.

\section{Results}
\label{Sec:Results}

The key question is as follows: at what point will an expensive SPS quality-assurance experiment have collected enough data to warrant shutting down?

\begin{table*}[htb!]
\centering
\caption{Optimised parameters with standard errors for the mean-rate function in Eq.~(\ref{Eq:FitNew}) when fitting $R(\tau; \theta) \times t$ to all the data available in a dataset.}
\label{Tab:BestFit}
\begin{tabular}{|c|c|c|c|c|c|}
\hline
Best Fit  & $g$                & $\tau_0$ (s)                   & $R_b$ (Hz)                       & $R_p$ (Hz)                      & $1/\gamma_p$ (s)               \\ \hline
1p2uW LS  & 0.23 ($\pm 11.72\%$) & \num{6.11e-8} ($\pm 0.05\%$) & \num{2.54e-4} ($\pm 24.37\%$) & \num{2.55e-3} ($\pm 1.20\%$) & \num{1.10e-9} ($\pm 13.20\%$) \\
1p2uW P   & 0.22 ($\pm 12.17\%$) & \num{6.11e-8} ($\pm 0.05\%$) & \num{2.66e-4} ($\pm 22.90\%$) & \num{2.59e-3} ($\pm 1.17\%$) & \num{1.05e-9} ($\pm 13.40\%$) \\ \hline
2p5uW LS  & 0.32 ($\pm 8.79\%$)  & \num{6.11e-8} ($\pm 0.06\%$) & \num{7.24e-4} ($\pm 39.84\%$) & \num{9.76e-3} ($\pm 1.37\%$) & \num{1.16e-9} ($\pm 15.13\%$) \\
2p5uW P   & 0.30 ($\pm 9.23\%$)  & \num{6.11e-8} ($\pm 0.06\%$) & \num{7.79e-4} ($\pm 36.30\%$) & \num{1.01e-2} ($\pm 1.32\%$) & \num{1.10e-9} ($\pm 15.33\%$) \\ \hline
4uW LS    & 0.36 ($\pm 7.31\%$)  & \num{6.11e-8} ($\pm 0.05\%$) & \num{9.13e-4} ($\pm 42.84\%$) & \num{1.30e-2} ($\pm 1.35\%$) & \num{1.18e-9} ($\pm 15.08\%$) \\
4uW P     & 0.34 ($\pm 8.01\%$)  & \num{6.11e-8} ($\pm 0.05\%$) & \num{1.09e-3} ($\pm 34.90\%$) & \num{1.34e-2} ($\pm 1.31\%$) & \num{1.07e-9} ($\pm 16.01\%$) \\ \hline
8uW LS    & 0.49 ($\pm 9.15\%$)  & \num{6.11e-8} ($\pm 0.10\%$) & \num{1.14e-3} ($\pm 79.85\%$) & \num{1.36e-2} ($\pm 2.75\%$) & \num{1.23e-9} ($\pm 32.09\%$) \\
8uW P     & 0.48 ($\pm 9.27\%$)  & \num{6.11e-8} ($\pm 0.09\%$) & \num{1.09e-3} ($\pm 81.37\%$) & \num{1.39e-2} ($\pm 2.66\%$) & \num{1.23e-9} ($\pm 30.62\%$) \\ \hline
10uW-- LS & 0.53 ($\pm 7.20\%$)  & \num{6.11e-8} ($\pm 0.08\%$) & \num{1.20e-3} ($\pm 91.99\%$) & \num{1.74e-2} ($\pm 2.54\%$) & \num{1.34e-9} ($\pm 27.54\%$) \\
10uW-- P  & 0.50 ($\pm 7.44\%$)  & \num{6.11e-8} ($\pm 0.08\%$) & \num{1.19e-3} ($\pm 87.34\%$) & \num{1.79e-2} ($\pm 2.37\%$) & \num{1.31e-9} ($\pm 26.15\%$) \\ \hline
10uW+ LS  & 0.51 ($\pm 5.92\%$)  & \num{6.11e-8} ($\pm 0.07\%$) & \num{4.46e-3} ($\pm 78.09\%$) & \num{7.33e-2} ($\pm 1.93\%$) & \num{1.26e-9} ($\pm 22.14\%$) \\
10uW+ P   & 0.50 ($\pm 6.15\%$)  & \num{6.11e-8} ($\pm 0.06\%$) & \num{4.92e-3} ($\pm 68.21\%$) & \num{7.48e-2} ($\pm 1.85\%$) & \num{1.20e-9} ($\pm 22.11\%$) \\ \hline
20uW LS   & 0.56 ($\pm 5.94\%$)  & \num{6.11e-8} ($\pm 0.07\%$) & \num{2.18e-3} ($\pm 49.65\%$) & \num{1.78e-2} ($\pm 2.39\%$) & \num{1.34e-9} ($\pm 26.62\%$) \\
20uW P    & 0.57 ($\pm 6.05\%$)  & \num{6.11e-8} ($\pm 0.07\%$) & \num{2.44e-3} ($\pm 44.00\%$) & \num{1.81e-2} ($\pm 2.34\%$) & \num{1.22e-9} ($\pm 28.61\%$) \\ \hline
30uW LS   & 0.68 ($\pm 5.97\%$)  & \num{6.11e-8} ($\pm 0.09\%$) & \num{2.62e-3} ($\pm 50.86\%$) & \num{1.34e-2} ($\pm 3.68\%$) & \num{1.39e-9} ($\pm 41.80\%$) \\
30uW P    & 0.68 ($\pm 5.96\%$)  & \num{6.11e-8} ($\pm 0.09\%$) & \num{2.76e-3} ($\pm 48.89\%$) & \num{1.36e-2} ($\pm 3.67\%$) & \num{1.31e-9} ($\pm 44.31\%$) \\ \hline
\end{tabular}
\end{table*}

If allowed to run for a seemingly long time, detecting $45000\pm25000$ coincidences, the experimental setups described by Table~\ref{Tab:Datasets} enable what will be called least-squares and Poisson-likelihood `best fits' for the parameters listed in Table~\ref{Tab:Bounds}.
These apparent ground truths are shown in Table~\ref{Tab:BestFit}.
Sure enough, as reported earlier~\cite{muzo20}, increasing the output power of the exciting laser diminishes the SPS quality of the studied InGaAs/GaAs epitaxial quantum dot.
Also of note, the fitting errors for $g$ are already not insubstantial.
While the location of the MPE peak and the maximum detection rate -- recall the comb structure in Fig.~\ref{Fig:ExampleHist} -- are easily identified, the background and the width of the peaks are harder to characterise with certainty.
But what of sampling-based errors?

For the present analysis, the generative method described in Sec.~\ref{Sec:Augmentation} works to `Poisson-sample' histograms representing the following numbers of total events: \num{1e3}, \num{1e4}, \num{1e5}, and \num{1e6}.
Naturally, every unique experiment is expected to accumulate these observations at different times.
For instance, based on the average event rate in Table~\ref{Tab:Datasets}, 1p2uW should encounter 10000 two-photon events at $t\simeq 3682$ s, while 10uW+ should observe 100000 at $t\simeq 2677$ s.
Nonetheless, given that a unit of information is more directly tied with an individual observation than the passage of a second, this dataset-dependent normalisation is essential for a fair comparison.

\begin{figure*}[htb!]
\centering
\begin{subfigure}{0.475\textwidth}
\centering
\includegraphics[width=\textwidth]{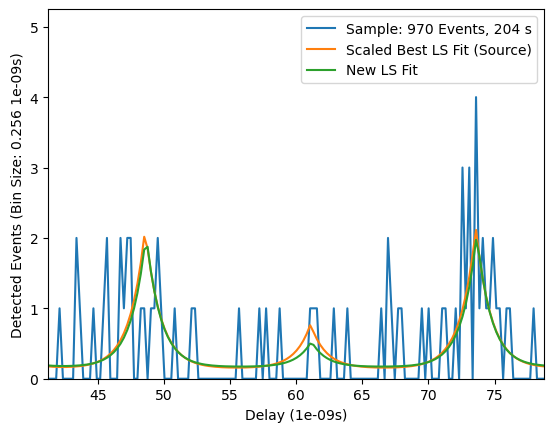}
\caption{}
\end{subfigure}
\hfill
\begin{subfigure}{0.475\textwidth}
\centering
\includegraphics[width=\textwidth]{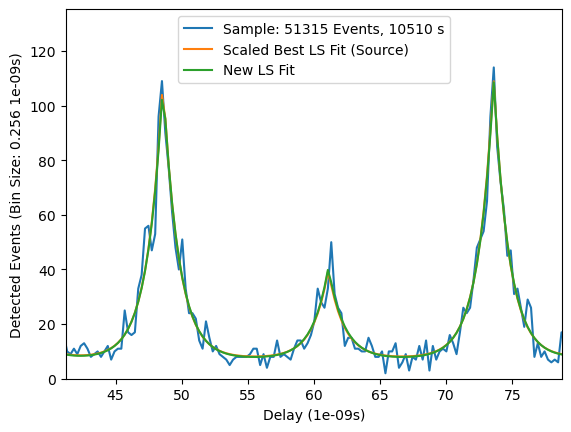}
\caption{}
\end{subfigure}
\begin{subfigure}{0.475\textwidth}
\centering
\includegraphics[width=\textwidth]{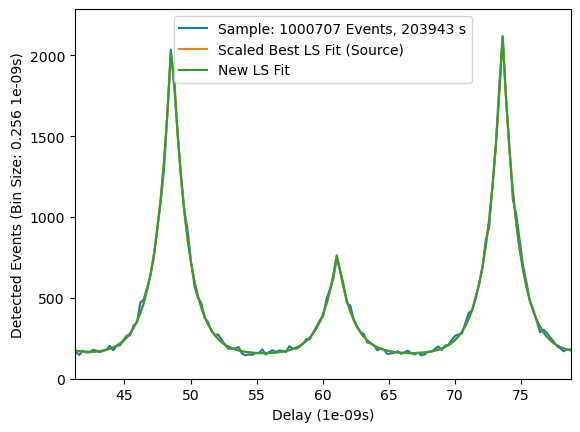}
\caption{}
\end{subfigure}
\caption{\label{Fig:MCExamples} Closeups of example histograms Poisson-sampled from the best least-squares fit for the 2p5uW experiment. Displays the original best fit from which the histogram is sampled as well as a new fit, both scaled for the appropriate duration $t$. Histogram contains approximately: (a) 1000 events, (b) 51534 events, i.e.~the original size of the full dataset that dictated the best fit, and (c) 1000000 events.}
\end{figure*}

\begin{figure}[htb!]
\centering
\includegraphics[width=\columnwidth]{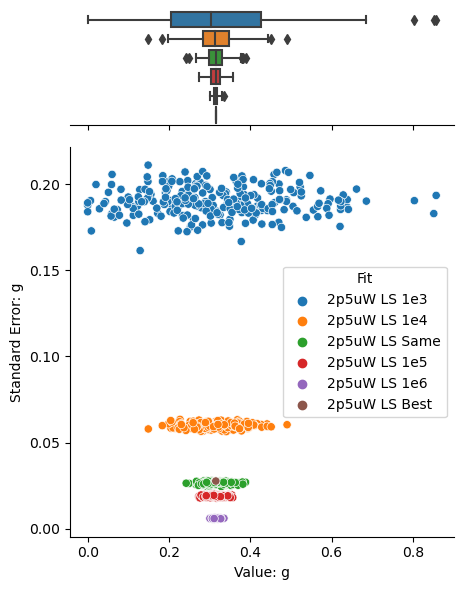}
\caption{\label{Fig:MCSpreadOneDataset} Comparison of least-squares fitted parameter $g$ between the `ground-truth' best fit for the 2p5uW experiment and fits applied to subsequent Poisson-sampled histograms. There are 250 histograms generated for each of the following approximate numbers of total events: \num{1e3}, \num{1e4}, \num{1e5}, and \num{1e6}. There are also 250 generated for approximately the same number of events as used for the best fit, i.e.~51534 for 2p5uW.}
\end{figure}

Accordingly, using the appropriate values of $t$, the next step is to optimise Eq.~(\ref{Eq:FitNew}) for any synthetic histograms generated.
Immediate inspection shows, as exemplified by Fig.~\ref{Fig:MCExamples} for the 2p5uW experiment and least-squares fitting methodology, that typical fits for sampled data become less and less visually discrepant with the appropriately scaled ground truth as more events are `observed'.
Of course, examining a single fit is of limited use; synthetically generated histograms prove most informative in bulk.
For instance, Fig.~\ref{Fig:MCSpreadOneDataset} shows how the least-squares fitted parameter $g$ varies across $250$ Poisson-samplings for each of the following sizes: ${\sim}1000$, ${\sim}10000$, ${\sim}100000$, and ${\sim}1000000$.
It also covers artificial histograms of the same size as the full experimental dataset, i.e.~${\sim}51534$ detections for 2p5uW.

\begin{figure}[htb!]
\centering
\includegraphics[width=\columnwidth]{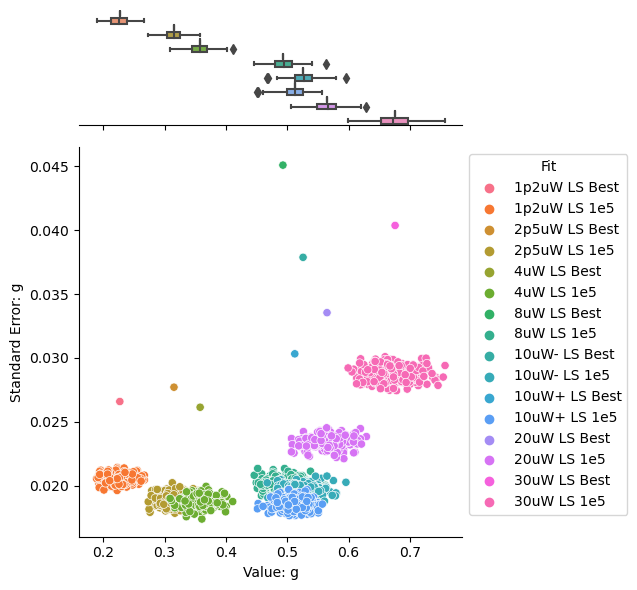}
\caption{\label{Fig:MCSpreadAllDatasets} Comparison of least-squares fitted parameter $g$, per dataset, between the `ground-truth' best fit and fits applied to 250 subsequent Poisson-sampled histograms of approximate size \num{1e5}.}
\end{figure}

\begin{table}[htb!]
\centering
\caption{Mean values and standard deviations for fitted parameter $g$, as derived from Poisson-sampled histograms of size \num{1e4} and \num{1e5} for different datasets. Both least-squares and Poisson-likelihood fits are represented. Also includes values for Poisson-sampled histograms containing the same number of events as each full dataset. Importantly, the sample-based standard deviation here is not the same as a fit-specific standard error.}
\label{Tab:MCSigma}
\begin{tabular}{|c|cc|cc|cc|}
\hline
\multirow{2}{*}{Best Fit} & \multicolumn{2}{c|}{\begin{tabular}[c]{@{}c@{}}$\num{1e4}$ Event\\ Histograms\end{tabular}} & \multicolumn{2}{c|}{\begin{tabular}[c]{@{}c@{}}Same Size\\ Histograms\end{tabular}} & \multicolumn{2}{c|}{\begin{tabular}[c]{@{}c@{}}$\num{1e5}$ Event\\ Histograms\end{tabular}} \\ \cline{2-7} 
                          & \multicolumn{1}{c|}{$\mu_g$}                          & $\sigma_g$                          & \multicolumn{1}{c|}{$\mu_g$}                      & $\sigma_g$                      & \multicolumn{1}{c|}{$\mu_g$}                          & $\sigma_g$                          \\ \hline
1p2uW LS                  & \multicolumn{1}{c|}{0.227}                            & 0.056                               & \multicolumn{1}{c|}{0.230}                        & 0.021                           & \multicolumn{1}{c|}{0.226}                            & 0.018                               \\
1p2uW P                   & \multicolumn{1}{c|}{0.218}                            & 0.059                               & \multicolumn{1}{c|}{0.219}                        & 0.022                           & \multicolumn{1}{c|}{0.211}                            & 0.017                               \\ \hline
2p5uW LS                  & \multicolumn{1}{c|}{0.315}                            & 0.054                               & \multicolumn{1}{c|}{0.317}                        & 0.025                           & \multicolumn{1}{c|}{0.315}                            & 0.017                               \\
2p5uW P                   & \multicolumn{1}{c|}{0.299}                            & 0.055                               & \multicolumn{1}{c|}{0.295}                        & 0.023                           & \multicolumn{1}{c|}{0.294}                            & 0.017                               \\ \hline
4uW LS                    & \multicolumn{1}{c|}{0.357}                            & 0.064                               & \multicolumn{1}{c|}{0.358}                        & 0.023                           & \multicolumn{1}{c|}{0.358}                            & 0.018                               \\
4uW P                     & \multicolumn{1}{c|}{0.333}                            & 0.056                               & \multicolumn{1}{c|}{0.332}                        & 0.023                           & \multicolumn{1}{c|}{0.332}                            & 0.019                               \\ \hline
8uW LS                    & \multicolumn{1}{c|}{0.486}                            & 0.072                               & \multicolumn{1}{c|}{0.492}                        & 0.052                           & \multicolumn{1}{c|}{0.494}                            & 0.021                               \\
8uW P                     & \multicolumn{1}{c|}{0.477}                            & 0.065                               & \multicolumn{1}{c|}{0.474}                        & 0.044                           & \multicolumn{1}{c|}{0.473}                            & 0.020                               \\ \hline
10uW-- LS                 & \multicolumn{1}{c|}{0.519}                            & 0.065                               & \multicolumn{1}{c|}{0.520}                        & 0.036                           & \multicolumn{1}{c|}{0.527}                            & 0.021                               \\
10uW-- P                  & \multicolumn{1}{c|}{0.492}                            & 0.060                               & \multicolumn{1}{c|}{0.494}                        & 0.039                           & \multicolumn{1}{c|}{0.494}                            & 0.020                               \\ \hline
10uW+ LS                  & \multicolumn{1}{c|}{0.511}                            & 0.073                               & \multicolumn{1}{c|}{0.507}                        & 0.034                           & \multicolumn{1}{c|}{0.512}                            & 0.020                               \\
10uW+ P                   & \multicolumn{1}{c|}{0.491}                            & 0.063                               & \multicolumn{1}{c|}{0.492}                        & 0.030                           & \multicolumn{1}{c|}{0.495}                            & 0.019                               \\ \hline
20uW LS                   & \multicolumn{1}{c|}{0.558}                            & 0.078                               & \multicolumn{1}{c|}{0.563}                        & 0.034                           & \multicolumn{1}{c|}{0.565}                            & 0.025                               \\
20uW P                    & \multicolumn{1}{c|}{0.567}                            & 0.074                               & \multicolumn{1}{c|}{0.566}                        & 0.035                           & \multicolumn{1}{c|}{0.567}                            & 0.026                               \\ \hline
30uW LS                   & \multicolumn{1}{c|}{0.664}                            & 0.100                               & \multicolumn{1}{c|}{0.674}                        & 0.045                           & \multicolumn{1}{c|}{0.674}                            & 0.032                               \\
30uW P                    & \multicolumn{1}{c|}{0.672}                            & 0.092                               & \multicolumn{1}{c|}{0.684}                        & 0.043                           & \multicolumn{1}{c|}{0.685}                            & 0.030                               \\ \hline
\end{tabular}
\end{table}

This result reveals an unfortunate implication.
Even after observing tens of thousands of two-photon events during the characterisation of an SPS under 2.5 \si{\uW} laser excitation power, the estimate of quality parameter $g$ will not just have a fitting-related uncertainty of ${\sim}0.025$; it could be off from a true characterisation by over ${\sim}0.1$, simply due to variability in Poisson processes.
Indeed, this sampling-revealed variance impacts all eight experiments, as Fig.~\ref{Fig:MCSpreadAllDatasets} demonstrates for histograms of size $\num{1e5}$.
That said, in fairness, the box plots in the figure indicate that most of the sampled fits are distributed much more tightly than they appear in the scatter plots.
Standard deviations are listed in Table~\ref{Tab:MCSigma}.
Accordingly, if there is a threshold for SPS quality, such as $g=0.5$, most of the listed experiments will be clearly classifiable even after observing only $10000$ events.
Nevertheless, in the case of 8 to 10 \si{\uW}, the stochastic nature of detection can easily swing determinations one way or the other.

\begin{figure}[htb!]
\centering
\includegraphics[width=\columnwidth]{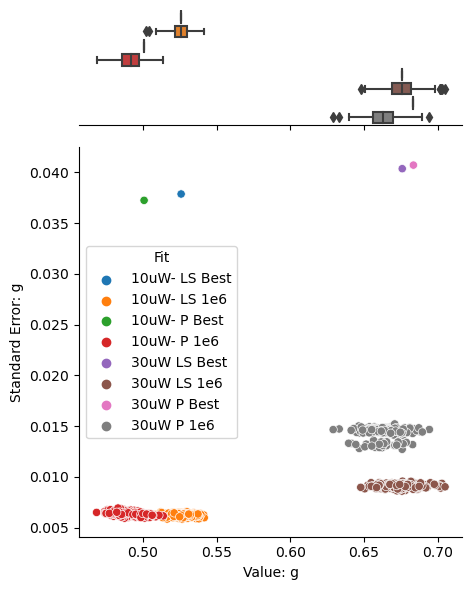}
\caption{\label{Fig:PIssue} Comparison of least-squares (LS) and Poisson-likelihood (P) optimisation for both the 10uW-- and 30uW experiments, juxtaposing the `ground-truth' best fits with those applied to 250 generated histograms of approximate size \num{1e6}. Specifically of note, above the scatter plot, there is misalignment between the best-fit `tick' and the corresponding box-plot mean for P fits.}
\end{figure}

Turning to the fitting methodologies described in Sec.~\ref{Sec:Fitting}, this work was unable to identify any significant or systematic benefit to optimising the Poisson-likelihood objective function in Eq.~(\ref{Eq:ObjectiveP}) proposed by previous research~\cite{coad20}.
Generally, the fitting and sample-based errors appear reasonably similar regardless of the method used.
Worse yet, while minimising the least-squares residual appears robust, almost universally aligning the mean value of fitted parameters for Poisson-sampled histograms with the original `best fit' value -- see the boxplots in Fig.~\ref{Fig:MCSpreadOneDataset} or Fig.~\ref{Fig:MCSpreadAllDatasets} -- this is not the case for the Poisson-likelihood method.
For instance, Fig.~\ref{Fig:PIssue} exemplifies cases where generated histograms should closely approximate the shape of the function they were Poisson-sampled from, being of size \num{1e6}, yet their P fits do not align with the best-fit parameter values.
It is unclear why this is; one hypothesis is that converging to an optimum for Eq.~(\ref{Eq:ObjectiveP}), where logarithms are involved, can face challenges of numerical stability.
More investigation is required to confirm this.
Nonetheless, as this work cannot validate a substantial comparative advantage of the MAP procedure, the rest of this article focusses solely on least-squares results.

Now, the takeaway thus far is necessarily cautioning.
The quality parameter $g$ is highly sensitive, crucially dependent on a small fraction of detected events across the $\tau$ delay domain and easily distorted by an incorrect background characterisation.
Hoping for a rapid and authoritative estimate of $g$ may be overly optimistic when even hours of accumulated observations leave a sizeable uncertainty.
Nonetheless, this work indicates how much uncertainty to expect at different two-photon event counts, providing, at a minimum, a `bootstrapping' procedure to apply in varying experimental contexts.
Since different practical applications involving single photons have varying requirements, one must decide what confidence level to embrace for a particular quantum optical measurement, subsequently accumulating detected events until the fitting and expected sampling errors are small enough to declare an SPS is `good' or `bad' for a particular purpose.

\begin{figure}[htb!]
\centering
\includegraphics[width=\columnwidth]{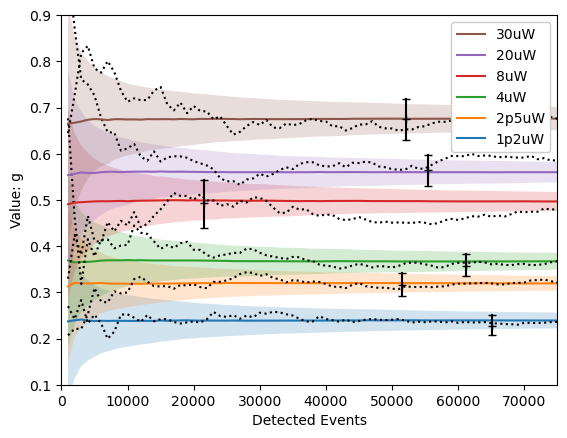}
\caption{\label{Fig:Expanding} The expected distributions for expanding averages of $g$, displayed for select datasets. Each expanding average is applied to a random iterative selection of histograms, where each histogram contains ${\sim}1000$ events; the dotted lines exemplify one such expanding average per dataset. Each shaded distribution is defined by the mean of $1500$ expanding averages, plus/minus one standard deviation. The black vertical `whiskers' mark the mean of $g$, plus/minus one standard deviation, derived from fitting histograms that contain as many events as each original dataset.}
\end{figure}

Granted, while small samples of two-photon events are highly unreliable and unlikely to carry discernible information about SPS quality, it is notable that averages over collections of small samples align well with individual fits over larger sample sizes.
The boxplots in Fig.~\ref{Fig:MCSpreadOneDataset} demonstrate this fact.
Admittedly, there is a lower limit for which this insight holds -- histograms of size $100$ are so sparse that there are systematic errors in fitting -- but sampling $1000$ events seems to meet this threshold decently.
One might then lean on an idea known well in machine learning and other fields, i.e.~ensembling weak estimators to create a singular powerful one.

Specifically, consider a steady accumulation of observed events in the vein of the eight datasets listed in Table~\ref{Tab:Datasets}.
After every $1000$ events, construct a histogram from only those $1000$ events, maintaining independent sampling, and then fit it.
Subsequently, average the parameters of that fit with those derived from all previous samples of size $1000$.
This concept is called an expanding average, and the value should generally improve over time.
Of course, it is still possible for this procedure to produce and sum long sequences of atypical histograms, e.g.~ones in Fig.~\ref{Fig:MCSpreadOneDataset} that are far from the mean, so a rigorous analysis requires the random but iterative Poisson-sampling of histograms, size $\num{1e3}$, to \textit{itself} be redone numerous times, e.g.~in the form of 1500 independent sequences generated for this work.
Given all this, the distributions of those 1500 expanding averages, plus and minus one standard deviation, are displayed in Fig.~\ref{Fig:Expanding} for several datasets.

These results show, especially considering the whiskers that denote the uncertainty in $g$ when fitting a single histogram of large size, that decent estimates of quality might be possible earlier on by averaging small-sample fits.
For instance, fitting the combined $52088$ events in the 30uW dataset provides a lower degree of confidence in $g$ than averaging the parameter over $52$ independent fits of $1000$ events.
This competitive advantage is true as early as $30000$ events.
Moreover, even where there is no clear-cut difference, the rapid tapering of confidence bounds still suggests that an expanding average can be a viable early estimator of SPS quality.
However, distributions provide no guarantee; the standard deviations of both expanding averages and large-sample fitting, depicted in Fig.~\ref{Fig:Expanding}, are probabilistic.
Some example trajectories, e.g.~for 8uW, veer substantially away from the distribution mean even at high numbers of samples.
Elsewhere, e.g.~for 1p2uW, the distribution itself does not align perfectly with the `best fit', possibly suggesting a minor systematic error in fitting for histograms of size $\num{1e3}$, just as is present for size $\num{1e2}$.
Essentially, the stochastic nature of two-photon event observations induces many complex challenges that stifle certainty for early estimates.

\begin{table}[htb!]
\centering
\caption{Background (BG) rates for the best least-squares fit applied to each full dataset. Each rate indicates the portion of events that are associated with BG detections, i.e.~a BG ratio.}
\label{Tab:BGRatio}
\begin{tabular}{|c|c|c|c|c|}
\hline
Dataset   & \begin{tabular}[c]{@{}c@{}}Best Fit\\ BG Rate $R_b$\\(Hz)\end{tabular} & BG Events & Total Events & \begin{tabular}[c]{@{}c@{}}BG\\ Ratio\end{tabular} \\ \hline
1p2uW LS  & \num{2.54e-4}                                       & 23787     & 65037        & 0.37                                                          \\ \hline
2p5uW LS  & \num{7.24e-4}                                       & 14868     & 51534        & 0.29                                                          \\ \hline
4uW LS    & \num{9.13e-4}                                       & 16748     & 61246        & 0.27                                                          \\ \hline
8uW LS    & \num{1.14e-3}                                       & 6507      & 21611        & 0.30                                                          \\ \hline
10uW-- LS & \num{1.20e-3}                                       & 7028      & 28817        & 0.24                                                          \\ \hline
10uW+ LS  & \num{4.46e-3}                                       & 10549     & 45198        & 0.23                                                          \\ \hline
20uW LS   & \num{2.18e-3}                                       & 20317     & 55469        & 0.37                                                          \\ \hline
30uW LS   & \num{2.62e-3}                                       & 24505     & 52088        & 0.47                                                          \\ \hline
\end{tabular}
\end{table}

\begin{figure*}[htb!]
\centering
\begin{subfigure}{0.475\textwidth}
\centering
\includegraphics[width=\textwidth]{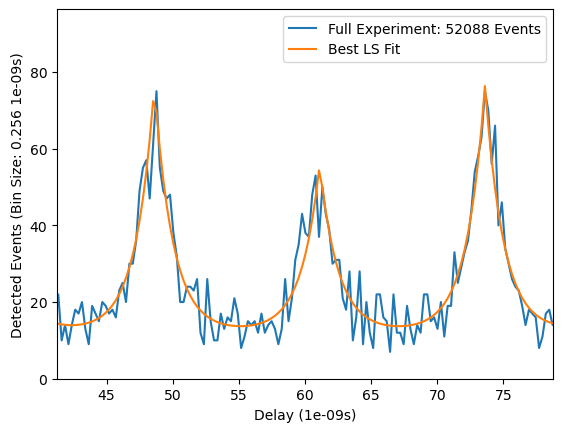}
\caption{}
\label{Fig:BG}
\end{subfigure}
\begin{subfigure}{0.475\textwidth}
\centering
\includegraphics[width=\textwidth]{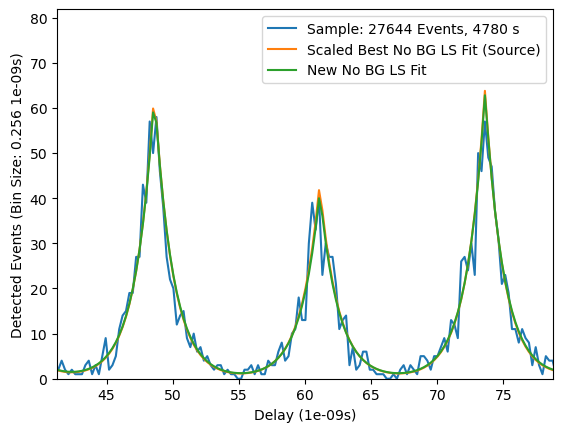}
\caption{}
\label{Fig:NoBG}
\end{subfigure}
\caption{Removing the two-photon event background for the 30uW dataset. (a) Closeup of the best fit over the full dataset. (b) A `same-duration' histogram Poisson-sampled from the best fit, but with $R_b$ set to zero. Note the decrease of about $12.5$ events per bin, i.e.~approximately $24505$ across $1954$ bins.}
\end{figure*}

In theory, some of this uncertainty could be better constrained if there was a way to identify and eliminate the background, i.e.~the model component in Eq.~(\ref{Eq:FitOld}) that is considered to be uniform across the delay domain.
After all, if the best-fit `ground truths' are to be believed, all the analysed datasets involve substantial counts of two-photon events, or at least triggered `detections', that are not due to SPS emission.
The ratios are listed in Table~\ref{Tab:BGRatio} and range from $0.23$ to $0.47$.
So, to investigate the effect of eliminating the background, the generative procedure in this work can be adjusted.
After acquiring a `best fit' of the five parameters in Table~\ref{Tab:BestFit}, visually depicted by Fig.~\ref{Fig:BG}, the background rate $R_b$ can be set to zero, subsequently allowing the generation of a `No BG' Poisson-sampled histogram, such as is exemplified in Fig.~\ref{Fig:NoBG}.

\begin{figure}[htb!]
\centering
\includegraphics[width=\columnwidth]{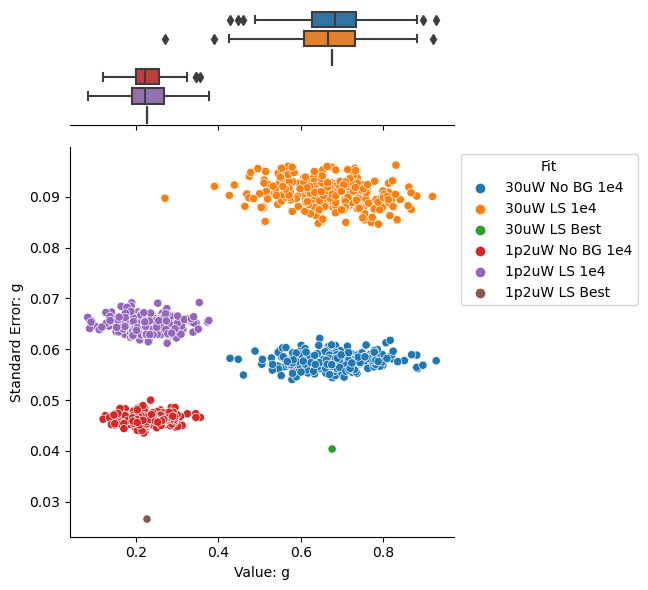}
\caption{\label{Fig:MCNoBG} Differences in fitted parameter $g$ between Poisson-sampled histograms that do and do not sample/fit background rate $R_b$. Note that `No BG' fits are applied to substantially less than $\num{1e4}$ events.}
\end{figure}

Crucially, without a background rate, the average event rate in Table~\ref{Tab:Datasets} and the total number of events in a sampled histogram are scaled by a factor of one minus the `BG Ratio' listed in Table~\ref{Tab:BGRatio}.
So, for instance, a 30uW `No BG' histogram of size $\num{1e4}$ actually contains around $5300$ events.
Even so, despite the substantially fewer events detected over the same measurement duration, Fig.~\ref{Fig:MCNoBG} shows that four-parameter fits applied to the `No BG' histograms -- $R_b$ in Table~\ref{Tab:Bounds} is set to zero and not allowed to vary -- considerably tighten their standard errors.
Background elimination is thus a worthwhile pursuit in acquiring better estimates of SPS quality.
However, as the figure demonstrates, this would not impact the variance intrinsic to the Poisson process that describes event detection; a significant degree of uncertainty is unavoidable.

\begin{figure}[htb!]
\centering
\includegraphics[width=\columnwidth]{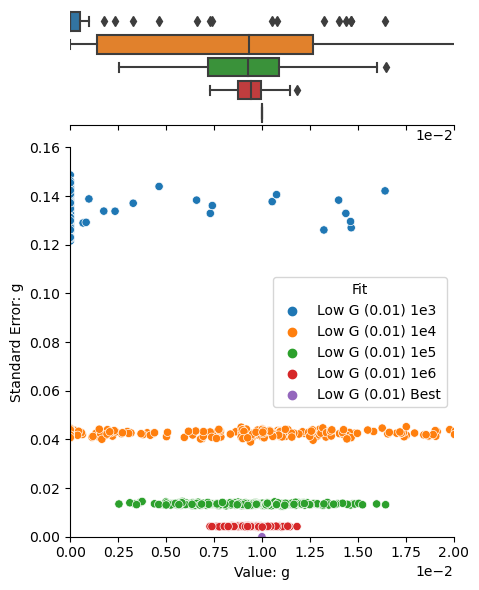}
\caption{\label{Fig:MCLowG} Demonstration of data augmentation applied in a low-$g$ scenario. A confected `ground-truth' histogram has best-fit parameters $g=\num{1e-2}$, $\tau_0=\num{6.11e-8}$, $R_b=0$, $R_p=\num{9.76e-3}$, and $1/\gamma_p=\num{1.16e-9}$. Subsequent fits are applied to 250 Poisson-sampled histograms per approximate size: \num{1e3}, \num{1e4}, \num{1e5}, and \num{1e6}. The resulting $g$ estimates are depicted, along with their least-squares fitting-based standard errors.}
\end{figure}

As a final reminder, the data augmentation technique in this paper is generally applicable, provided that certain physical assumptions hold.
Accordingly, consider a state-of-the-art low-$g$ SPS device~\cite{mita16, hafi18, scjo18} with the following ground-truth emission profile: $g=\num{1e-2}$, $\tau_0=\num{6.11e-8}$, $R_b=0$, $R_p=\num{9.76e-3}$, and $1/\gamma_p=\num{1.16e-9}$.
As Fig.~\ref{Fig:MCLowG} shows, one can Poisson-sample histograms to quantify uncertainty for this device in much the same manner as for the
FI-SEQUR quantum dot.
However, there are caveats.
With such a tiny MPE peak, sufficient detections must be gathered to even acknowledge its existence.
For example, fitting synthetic histograms of \num{1e3} events predominantly and falsely suggests a perfect $g=0$; any exceptions are statistical outliers.
Moreover, even when the number of two-photon events is sufficiently large to resolve $g$, e.g.~when the $250$ histograms of size \num{1e6} statistically indicate $g=\num{9.368e-3}\pm\num{0.847e-3}$, one must also consider whether the standard error of the fitting algorithm itself washes out the $g$ value of interest.
The unfortunate takeaway is that acquiring sufficient confidence in more extreme values of $g$ requires a much greater signal-to-noise ratio, e.g.~by increasing CPS or integration times.
Whatever the case, the power of the data augmentation technique is that it can quantify these expectations for experimentalists.

\section{Discussion}
\label{Sec:Discussion}

One of the primary motivators of this work is the publication of a MAP method for SPS quality estimation~\cite{coad20} that promised up to two orders of magnitude shorter data acquisition times. Such a claim is understandably alluring to experimentalists, as performing statistical measurements on quantum emitters is resource expensive and time consuming.

Unfortunately, this statement must be interpreted carefully.
First of all, experimental context matters. For instance, the laser pulses in the MAP publication~\cite{coad20} appear to have a period of $1$ \si{\us}, as opposed to the $12.5$ ns in this work.
This detail means that sequential peaks in the data are well separated and distinct; inaccuracies in fitting the decay factor $\gamma_p$ are unlikely to affect confidence in $g$ as much as they do for the compressed comb structures in this work.

Admittedly, a reduced frequency of excitation pulses does decrease the number of photons detected within a delay window of interest, so there is a trade-off.
Indeed, in the case of many practical applications, it is appealing to drive the SPS as often as possible to maximise the single-photon generation rate.
Hence, assessing SPS quality under regimes of higher-frequency excitation retains its appeal.
Of course, there is a fundamental limit here, in that the SPS cannot be excited until it has time to relax; the spontaneous-emission lifetime for the FI-SEQUR system is around one to two nanoseconds, and the peaks cannot be brought closer to each other than that.
However, if decay factor $\gamma_p$ could be confidently fitted by a quick supplementary experiment, e.g.~involving distinct excitation pulses, then perhaps overlapping peaks would no longer be a major factor of uncertainty, enabling many more valuable observations in the same amount of time without any of the drawbacks.

Notably, time is a poor proxy for the pace of accumulating information if the rates of detecting two-photon events vary dramatically between experiments. A brief measurement can correspond to many two-photon events if the count rates are high.
Such a reason is why this work aligned synthetic histograms across datasets by the number of events observed rather than the time passed.
Indeed, for the sake of fair comparison between different methods of SPS quality estimation, one has to consider histograms of equal information content: an equivalent number of events observed across the given time domains.
If so, maybe being able to sample so many repeated peaks proves minorly beneficial when fitting FI-SEQUR data, in that $R_p$ becomes accurately characterised.
However, since $g$ relies heavily on the MPE peak, early quality estimates may actually benefit more from a relatively high proportion of observations existing in that central zone.
Essentially, the $50$-second integration time presented in the MAP article~\cite{coad20} may provide many more \textit{informative} events within the relevant domain than most FI-SEQUR datasets do for the same amount of time; see Fig.~\ref{Fig:ExampleHistCloseup} and extrapolate from the ten-second histogram.

The signal-to-background ratio is another complicating factor.
Table~\ref{Tab:BGRatio} in this work shows that the FI-SEQUR datasets contain numerous `false' detection events that, if eliminated, would significantly improve estimates; see Fig.~\ref{Fig:MCNoBG}. In contrast, the experimental data considered in the MAP publication~\cite{coad20} appears to have no background.
Of course, the feasibility of independent background characterisation within an experimental context, let alone its elimination, is another matter to discuss elsewhere.
The point here is that, second for second, the correlation measurement setup in the MAP paper appears to observe many more two-photon events under adjacent peaks than the FI-SEQUR experiments detailed here.

Essentially, this work was unable to confirm the suggested advantage of the Poisson-likelihood approach~\cite{coad20} compared to least-squares fitting.
Uncertainties in $g$, due to both fitting errors and stochastic effects, remain unavoidable for now.
Of course, if the claim was instead referring to the efficiency of computationally estimating $g$, regardless of its error, then the reformulation provided by Eq.~(\ref{Eq:FitNew}) of this work will find appreciation for its added algorithmic utility.

\section{Conclusion}
\label{Sec:Conclusion}

Determining the second-order correlation function that describes the emission statistics of a light source requires the acquisition of two-photon events.
In the context of intended SPS quantum dots, such measurements are costly in both time and material resources.
Consequently, there exists a research endeavour focussed on estimating SPS quality as quickly and accurately as possible.
However, this work cautions that recent optimism in the literature, promoting new techniques capable of ``fast estimates in under a minute'' and ``a one-to-two order of magnitude speed-up'', may need to be tempered.

A central concern is that novel estimation approaches may be promoted without validation on a sufficient amount of data.
One singular experiment, i.e.~one instance of how a histogram binning observed events evolves, is simply not enough.
Hence, assuming detections adhere to Poisson statistics, this work centres on a generative method similar to bootstrapping that allows for the computational synthesis/analysis of new data from existing datasets.
Specifically, having applied this process of data augmentation to eight datasets studying a fibre-coupled InGaAs/GaAs epitaxial quantum dot, this work assesses estimates of the SPS figure of merit $g$, a.k.a.~$g^{(2)}(0)$, and discusses the impact of stochastic variability in the measurements.

The major contributions of this work are as follows:
\begin{itemize}
\item The reformulation of well-known theoretical equations describing the expected delay-domain histograms of detected two-photon coincidences, in the context of an SPS excited by a pulsed laser, thus allowing data to be fit in a computationally efficient manner.

\item Proof of principle that data augmentation is a valuable tool in quantifying the unavoidable systematic errors in $g$ that result from the variability of Poisson processes. This work cautions that neglecting these errors, a common oversight in the literature, can lead to unwarranted overconfidence in early $g$ estimates and premature declarations of SPS devices as state-of-the-art.

\item A comparison between the standard least-squares approach and a recently proposed Poisson-likelihood method, finding no significant/systematic advantage for the latter and even a potential vulnerability to numerical instabilities during optimisation.

\item An investigation of whether expanding averages over small-size histograms is competitive against $g$ estimation based on a single large-size histogram.
The results suggest expanding averages show some promise for early estimation, but deeper analysis is required to determine their general utility rigorously.

\item The finding that suppressing background counts would boost the fitting accuracy of $g$ but has no significant discernible effect on the error from the stochastic variability in detections.
\end{itemize}

Ultimately, this research serves as another example of how experimental physics can benefit from applying data-driven approaches commonly found in machine learning, such as data augmentation.
In this case, provided the real-world data adheres well to the assumptions underlying a theoretical model, experimental observations can be supplemented with numerical `extrapolations' from which bootstrapped statistics can be deduced.
Only with this improved understanding can methods for assessing SPS quality be judged appropriately for speed \textit{and} accuracy.

\begin{acknowledgments}
The experimental data used in this study was obtained within the FI-SEQUR project, which was jointly financed by (1) the European Regional Development Fund (EFRE) of the European Union, as part of a programme to promote research, innovation, and technology (Pro FIT) in Germany, and (2) the National Centre for Research and Development in Poland, as part of the 2nd Poland-Berlin Photonics Programme, grant number 2/POLBER-2/2016 (project value: 2089498 PLN). 
\end{acknowledgments}

\section*{Author Declarations}
\subsection*{Conflict of Interest}
The authors have no conflicts to disclose.

\subsection*{Author Contributions}
\textbf{David Jacob Kedziora}: Conceptualisation (supporting); Formal Analysis (lead); Methodology (lead); Software (lead); Visualisation (lead); Writing -- original draft (lead); Writing -- review \& editing (equal). \textbf{Anna Musiał}: Conceptualisation (equal); Data Curation (lead); Investigation (equal); Writing -- review \& editing (equal). \textbf{Wojciech Rudno-Rudziński}: Investigation (equal); Writing -- review \& editing (equal). \textbf{Bogdan Gabrys}: Conceptualisation (equal); Methodology (supporting); Project Administration (lead); Supervision (lead); Writing -- review \& editing (equal).

\section*{Data Availability}
The data that supports the findings of this study is openly available at https://github.com/UTS-CASLab/sps-quality.

\bibliography{sps}

\end{document}